\documentclass[10pt,twocolumn,preprintnumbers,amsmath,amssymb,nofootinbib
,superscriptaddress]{revtex4}
\usepackage{graphicx,longtable,mathrsfs,color,array}
\usepackage[hidelinks]{hyperref}
\usepackage[usenames,dvipsnames]{xcolor} % colour
\usepackage{amssymb,amsmath,mathtools,mathrsfs,slashed} % maths
\usepackage{epsfig,subfigure,placeins,float} % plots
\usepackage{booktabs,longtable,ctable,multirow} % tables
\usepackage{exscale,relsize} % scale 
\usepackage[normalem]{ulem} % editing
\usepackage{enumerate}
\usepackage{times, mathptmx} % Times font
\usepackage{color}
\usepackage{hyperref}
\usepackage{graphicx}% Include figure files
\usepackage{color}
\usepackage{graphicx,graphics}

\newcommand{\be}{\begin{equation}}
\newcommand{\ee}{\end{equation}}
\newcommand{\bea}{\begin{eqnarray}}
\newcommand{\eea}{\end{eqnarray}}

\usepackage[stable]{footmisc}
\usepackage{color}

\def\dkmu2{\delta K_{\mu \nu}\delta K^{\mu \nu}}
\def\pmu2{  \phi_{\mu \nu}\phi^{\mu \nu}}

\begin{document}

\title{No fifth force in a scale invariant universe.}

\author{Pedro G. Ferreira}
\email{pedro.ferreira@physics.ox.ac.uk}
\affiliation{Astrophysics, University of Oxford, DWB, Keble Road, Oxford OX1 3RH, UK}
\author{Christopher T. Hill}
%\email{p.j.bull@astro.uio.no}
\affiliation{Fermi National Accelerator Laboratory, P.O.Box 500, Batavia, Illinois 60510, USA}
\author{Graham G. Ross}
%\email{p.ferreira1@physics.ox.ac.uk}
\affiliation{Theoretical Physics, University of Oxford, 1 Keble Road, Oxford OX1 3NP, UK}
\date{Received \today; published -- 00, 0000}

\begin{abstract}
We revisit the possibility that the Planck mass is { spontaneously generated in 
scale invariant scalar-tensor theories of gravity,  typically leading to a ``dilaton.''} 
The fifth force, arising 
from the {dilaton}, is severely constrained by astrophysical measurements.  
We explore the possibility that nature is fundamentally {Weyl-scale} invariant and argue that, 
as a consequence, the  fifth force effects are dramatically suppressed and such
models are viable.
We discuss possible obstructions to {maintaining} scale invariance and how these might be resolved.
\end{abstract}

%\pacs{98.80.-k, 98.80.Es, 95.36.+d, 95.36.+x}

\maketitle

\section{Introduction}
\label{sec:intro}
The possibility that the gravitational constant, $G$, or alternatively the Planck mass, $M_{\rm Pl}$ 
is {dynamically generated} has been considered for more than half a century. 
P.~Dirac argued that the large number hypothesis indicated the possibility 
that $G$ obeyed an equation of the form $\Box G \sim \rho$ while C.~Brans and R.~Dicke 
proposed the action
\begin{eqnarray}
S_{BD}=
-\int d^4 x\sqrt{-g}\left[-\frac{\alpha}{12}\phi^2R
+\frac{1}{2}g^{\mu\nu}\partial_\mu\phi\partial_\nu\phi-V(\phi)
+L_m\right] \nonumber \\ 
\label{JBD}
\end{eqnarray}
where $g_{\mu\nu}$ is the metric, $R$ the corresponding 
Ricci scalar, $L_m$ is the matter Lagrangian, minimally coupled to 
$g_{\mu\nu}$ and $\alpha$ is a dimensionless constant \cite{Brans:1961sx} 
(we are assuming the mostly minus sign convention). Brans and Dicke's original 
theory is normally expressed in terms of the dynamical Planck mass 
$\Phi=-\frac{\alpha}{6}\phi^2$, $V=0$ and the parameter $\omega_{\rm BD}\sim1/\alpha$. 
{The} Brans-Dicke action has become one of the workhorses of gravitational physics and is 
used to explore extensions of general relativity that appear in a wide range of 
fundamental contexts. It has a more modern, complete incarnation - the Horndeski action - which 
encapsulates all possible Scalar-Tensor theories  which have second order equations 
of motion \cite{Horndeski:1974wa}.

If scalar tensor theories are to work, we {require} a mechanism by which 
the Planck mass stabilizes at its observed value. This can be achieved through a variety of ways, 
most notable by picking a potential $V$ such that, during its cosmic history, 
the scalar field settles at its minimum. The potential can have an explicit mass 
scale, {$\phi_0$}, in it (e.g. of the form $V\sim (\phi^2-\phi^2_0)^n$ where $n$ 
can be positive or negative). The curvature 
of the effective potential will then set the effective range of the fifth force 
arising from the scalar field; a judicious choice of curvature can lead to a small 
enough range that current observational constraints can be avoided.

The non-minimal coupling of $\phi$ with $R$ can lead to a richer variety of dynamics than those observed for standard scalar fields. In particular it is possible to construct models such that there are no dimensionfull parameters. If we choose $V=\lambda \phi^4$ and observe that, in the absence of matter, the equation of motion for $\phi$ can be cast as
\begin{eqnarray}
(1-\alpha)\left[\Box\phi+\frac{\nabla_\mu\phi\nabla^\mu\phi}{\phi}\right]+\phi^4\frac{d}{d\phi}\left(\frac{V}{\phi^4}\right)
\end{eqnarray} 
we find that the homogenous solution satisfies
\begin{eqnarray}
a^3\phi\frac{d\phi}{dt}= {\rm constant}
\end{eqnarray}
where $a$ is the scale factor of the Universe. 
In an expanding Universe, ${\dot \phi}\rightarrow 0$ and $\phi\rightarrow \phi_0$. 
The final value will not be set by the minimum of the potential but by the field's initial value. 
{
This is a universe of eternal inflation and a spontaneously generated Planck mass,
as described in the single scalar model of \cite{Ferreira:2016wem}. It is seen to be equivalent 
(in the Einstein frame) to a theory with a cosmological constant, fixed Planck scale 
and a completely decoupled dilaton. In two-field, or more, generalizations we can have inflation, 
Planck scale generation, and end up in a vacuum with vanishing cosmological constant.
 The time evolution naturally evolves the system from a Jordan
frame to an Einstein frame \cite{Ferreira:2016wem}.}

{Generic} Scalar-Tensor theories are very severely constrained by observations. 
The process of estimating these constraints is well established (and clearly presented in the original 
paper by Brans and Dicke \cite{Brans:1961sx} and then generalized in \cite{1992CQGra...9.2093D} 
and \cite{2014grav.book.....P}). While the original calculation \cite{Brans:1961sx} was done in 
the Jordan frame (i.e. the frame in which $\phi$ is non-minimally coupled to $R$) it has 
now become customary to  transform to the Einstein frame where we have the standard 
Einstein-Hilbert action but where $L_m$ is now coupled to $A(\phi)g_{\mu\nu}$ (where $A(\phi)$ 
arises from the conformal transformation between frames). The direct coupling between $\phi$ 
and matter brings out the interpretation of $\phi$ as the mediator of a {``fifth force''} which 
supplements the ordinary gravitational force.

The presence of $\phi$ {leads} to modifications of the usual solutions to the Einstein equations. 
For example the Schwarzschild-like solutions will have two non-trivial Parametrized Post Newtonian 
(PPN) parameters, $\gamma$ and $\beta$, which can be constrained using, for example, measurements of the 
Shapiro time delay, light deflection and the Nordvedt effect. The current, tightest constraints  come from 
an analysis of the Cassini spacecraft placing an upper bound on $\gamma$  such that 
$\omega_{\rm BD}> 40,000$ \cite{Bertotti:2003rm}. Comparable constraints 
(within a factor of $2$) have been obtained from the analysis of the relativistic 
pulsar-white dwarf binary, J1738+033 \cite{2012MNRAS.423.3328F}.

In this letter we will show that these results can be evaded if $L_m$ is {\it scale invariant} 
\footnote{We will in fact use Weyl invariance, {which is a multiplicative scale transformation of 
fields incuding the metric that does not affect coordinates}, 
as the key property. { Diffeomorphism scale invariance, which transforms the coordinates, 
$\delta x^\mu = \epsilon(x)x^\mu$ and is more general,
inevitably arises from the fact that we are considering generally covariant theories without mass
scales}}. 
{That is}, when we transform $g_{\mu\nu}\rightarrow \Omega^{-2}g_{\mu\nu}$, 
$\phi\rightarrow \Omega\phi$  and the matter appropriately (where $\Omega$ is a {\it constant}), 
we find that $S_M=\int d^4 x\sqrt{-g}L_m$ is invariant {and} then the fifth force is completely non-existent. 
That the fifth force should be suppressed  is simple and in some sense, obvious. 
In this scenario, scale invariance is a global {Weyl} symmetry which is broken when the scalar 
field settles down to its asymptotic value, {i.e.} when the Planck mass is stabilised. 
As a result, there is a Goldstone boson - the dilaton - which is the mediator of the 
fifth force and is, at most, derivatively coupled to the matter sector. 

Derivative couplings between the dilaton and the matter sector will lead to a 
suppression, {at large distance in the 5th force}. We will show, in detail,  that, in fact, in a 
scale invariant universe, in the symmetry broken phase, the dilaton does not perturbatively couple 
{\it at all} to the matter energy momentum tensor. There are {several} ingredients 
to this effect. The first one is that, given that the dilaton is derivatively coupled, 
the relevant terms  in the action are of the form
$[\delta {L}/\delta(\nabla^\mu\sigma)]\nabla^\mu\sigma=K_\mu\nabla^\mu\sigma $ where $L$ 
is the Lagrangian including all terms in $\sigma$ and $K_\mu$ is a four current, {the global
Weyl symmetry current,
%as we will show, 
which is conserved. In the symmetry broken phase, $K_\mu=0$ and thus 
$\sigma$ decouples from the matter dynamics of the theory}.
%in the symmetry broken phase.
The second ingredient is that the kinetic term 
for fermions involves a symmetrized derivative,
$({\overrightarrow {\slashed{\nabla}}}-{\overleftarrow {\slashed{\nabla}}})$, 
which is completely blind to a {\it real} rescaling of the fermion field (as opposed to one 
involving a complex phase). {Thirdly, gauge fields are neutral
under Weyl transformations, and the dilaton is automatically decoupled
from a classical gauge action.  
If the Weyl symmetry is valid at the quantum level as well,
again the dilaton completely decouples from the gauge field action}. 
We will use this paper to show how all these ingredients come into play 
and flesh out the proof that scale invariant theories evade fifth constraints by examining a 
number of specific examples. 

We begin, in section Section \ref{sec:standard} by re-deriving the  
constraints in the standard derivation and {then} show how 
they may be evaded for the case of single scalar field coupled 
to a scale invariant matter action - we should expect, at most, 
a derivative coupling of the dilaton to the matter source. 
In the Section \ref{sec:multi} we delve deeper and explored 
how the Weyl symmetry of a multi-scalar field action actually 
leads the dilaton to completely decouple from the matter sector.  
In Section \ref{sec:SM} we build on our previous results and show how 
a proxy for the standard model - a fermion that acquires a mass via the Higgs mechanism - 
will lead to the same result. We also demonstrate the decoupling of the dilaton from gauge bosons. Finally we discuss non-perturbative effects that can lead to the coupling of the dilaton to matter, but in a highly suppressed manner.

{ In Section \ref{sec:obstructions} we briefly address the 
fact that, while quantum corrections would seem
to invalidate scale invariance via trace anomalies,  this problem can be avoided
\cite{Ferreira:2016wem}.
It is well known that a quantum theory in $D=4$ with no input masses and vanishing $\beta$-functions
to all orders in $\hbar$ is scale symmetric.
However, the converse is not necessarily true:
a quantum theory in $D=4$ with no input masses and non-vanishing $\beta$-functions
is not necessarily non-scale invariant. That is, it can be interpreted as a subsector of a 
fully scale invariant theory. In these theories, ratios of observables to fixed mass scales, such as
$\phi_c/M$,  where  $\phi_c$ is a classical field VEV (or an
external momentum scale in a scattering amplitude) and $M$ is a fixed mass scale, 
do not occur as arguments of logs. Rather renormalization group running occurs in Weyl
invariant ratios, e.g., $\phi_c/\chi_c$
which respects overall scale symmetry \cite{Ferreira:2016wem}. In short, in these theories
there is no absolute mass scale
in nature, but rather just dimensionless ratios of VEV's.}
In Section \ref{sec:discussion} we {summarize} our findings.

%%THE STANDARD CALC
\section{Evading fifth force constraints with the dilaton.}
\label{sec:standard}
We take as our starting point the action presented in Equation \ref{JBD}. The modified Einstein field equations are
\begin{eqnarray}
-M^2G_{\alpha\beta}=& &\left(1-\frac{\alpha}{3}\right)\partial_\alpha\phi\partial_\beta\phi-\left(\frac{1}{2}-\frac{\alpha}{3}\right)\partial_\mu\phi\partial^\mu\phi g_{\alpha\beta}\nonumber \\&+&\frac{\alpha}{3}\left(\phi\box\phi g_{\alpha\beta}-\nabla_\alpha\nabla_\beta\phi\right)+Vg_{\alpha\beta}-T_{m\alpha\beta} \label{EFEs}
\end{eqnarray}
where we have defined $M^2=-\alpha\phi^2/6$. The modified Klein-Gordon equation is
\begin{eqnarray}
\Box\phi+\frac{\alpha}{6}R\phi+V_\phi=0 \label{KGs}
\end{eqnarray}
where $V_\phi=dV/d\phi$.

We are interested in studying these equations in two limits. First we will expand around Minkowski space, $\eta_{\alpha\beta}$ and will assume that $\phi$ has stablized around a minimum value, $\phi_0$. Hence we are interested in linear fluctuations around the scalar field minimum, $\phi=\phi_0+\varphi$ and the Minkowski metric,
$g_{\alpha\beta}=\eta_{\alpha\beta}+{\rm diag}(\Phi,\Psi\delta_{ij})$. Second, we are interested in the Newtonian, or quasi-static regime where we can discard all time derivatives of the metric and scalar field.  Taking the trace of Equation \ref{EFEs} to eliminate the Ricci scalar in Equation \ref{KGs} and the taking the two approximations described above we end up with
\begin{eqnarray}
\nabla^2\varphi=-\frac{\alpha}{6(1-\alpha)}\frac{\phi_0}{M^2_{\rm Pl}}T_m \label{varphiQS}
\end{eqnarray}
where we have defined $M^2_{\rm Pl}\equiv-\alpha\phi_0^2/6$ and we have assumed that contributions from $d^2V/d\phi^2$ are negligible. 

The Einstein field equations become, in terms of the gravitational potentials,
\begin{eqnarray}
-M^2_{\rm Pl}\nabla^2\Psi&=&-\frac{1}{2}\frac{3-2\alpha}{3(1-\alpha)}T_{m00} \nonumber \\
-M^2_{\rm Pl}\nabla^2(\Phi-\Psi)&=&\frac{2\alpha}{3(1-\alpha)}T_{m00}
\end{eqnarray}
where we have assumed a non-relativistic source, $T_m\simeq T_{m00}$ and $T_{mij}\simeq 0$. A localized mass gives us
$T_{m00}\simeq M\delta^3({\bf r})$ and we can solve for the potentials to find
\begin{eqnarray}
\Psi&=&-\frac{3-2\alpha}{6(1-\alpha)}\frac{1}{M^2_{\rm Pl}}\frac{M}{r} \nonumber \\
\Phi&=&\frac{1}{2(1-\alpha)}\frac{1}{M^2_{\rm Pl}}\frac{M}{r}
\end{eqnarray}
If we define Netwon's constant via $\Phi=G_0M/r$ we have that the PPN parameter $\gamma$ defined through
\begin{eqnarray}
\Psi\equiv \gamma \frac{G_0 M}{r} \nonumber
\end{eqnarray}
is given by
\begin{eqnarray}
\gamma=\frac{2\alpha-3}{4\alpha-3}. 
\end{eqnarray}
We have recovered the well established expression for $\gamma$ for scalar-tensor theories.

Crucial, in this derivation, is the fact that $\varphi$ is sourced by $T_m$ and furthermore, that the
energy momentum tensor of $\varphi$ then sources the gravitational potentials. Because of the non-minimal
coupling, $\varphi$ enters the Einstein field equations in combinations of the form $\phi_0\nabla^2\varphi$, bringing in extra contributions of $T_m$ to the right hand side. We can immediately see that, if the energy momentum tensor of matter fields is traceless there is no extra contribution to the metric potentials.  

If the action presented in Equation 
\ref{JBD} is scale invariant, the situation changes dramatically. 
Specifically, assume that $V=\frac{\lambda}{4}\phi^4$ and that, under 
scale transformations, $\sqrt{-g}L_m$ is invariant. Then consider the 
following {Weyl} field redefinitions
\begin{eqnarray}
\phi&=&\phi_0 e^{\frac{\sigma}{f}} \nonumber \\
g_{\alpha\beta}&=&{\hat g}_{\alpha\beta}e^{-\frac{2\sigma}{f}}
\end{eqnarray}
where $\phi_0$ is the stationary solution of the background field equations and $\sigma$ is a scalar field - the dilaton.  Transforming the action,
we find
\begin{eqnarray}
S_{BD}\rightarrow\int d^4 x\sqrt{-{\hat g}}\left[-\frac{\alpha}{12}\phi_0^2{\hat R}+\frac{1}{2}{\hat g}^{\mu\nu}\partial_\mu\sigma\partial_\nu\sigma-V(\phi_0)+{\hat L_m}\right] \nonumber \\ \label{JBD1}
\end{eqnarray}
where we have chosen $f=(1-\alpha)\phi^2_0$ so as to canonically normalize $\sigma$. Note that, because of our assumptions about scale invariance, the transformed matter action, ${\hat L}_m$ does not couple directly to the dilaton $\sigma$ although it may, however, couple to $\partial_\alpha\sigma$. This means that the dilaton equation of motion will be of the form
\begin{eqnarray}
\Box\sigma=\partial_\alpha\sigma S^\alpha
\end{eqnarray}
where $S^\alpha$ is constructed from elements of $T_{m\mu\nu}$. In fact, it is likely that $S^\alpha=\partial^\alpha S$ where $S$ is a local function of the matter fields. We then have that $\sigma$ is non zero inside the source but  satisfies $\Box\sigma=0$ outside, i.e. a damped wave equation. This means that, at late times, any  contribution from $\sigma$ to the energy momentum tensor sourcing the Einstein field equations is severely surpressed (as we shall see in Section \ref{halos}) and the standard constraints on Jordan-Brans-Dicke gravity do not apply. 

A key aspect to this derivation is the scale invariance of $L_m$. We have assumed that there will be a derivative coupling to $\sigma$ as we would expect from Goldstone's theorem. For this coupling to be completely absent, as we saw above, we would naively expect that  we would have to restrict ourselves to a conformally matter source and that the result, therefore, follows trivially from our original derivation. In the next section we will dig a bit deeper and consider explicit forms for $L_m$ to see that this is not necessarily the case.

%%MULTIFIELDS
\section{The dilaton in a multi-scalar universe.}
\label{sec:multi}
Let us now consider a multi-scalar tensor theory of gravity of the form
\begin{eqnarray}
S=\int d^4 x\sqrt{-g}\left[-\frac{1}{12}\sum_{i}^{N}\alpha_i\phi_i^2R+\frac{1}{2}\sum_{i}^{N}\partial_\mu\phi_i\partial^\mu\phi_i -W({\vec \phi})\right] \nonumber \\ \label{msaction}
\end{eqnarray}
where we assume a generalized "$\lambda\phi^2$ potential of the form:
\begin{eqnarray}
W({\vec \phi})=\sum_{i}^N\sum_{i}^N \phi^2_i W_{ij}\phi^2_j \nonumber
\end{eqnarray}
The action in Equation \ref{msaction} is scale invariant: it is invariant under $g_{\mu\nu}\rightarrow \Omega^{-2}g_{\mu\nu}$, $\phi_i\rightarrow \Omega\phi_i$ where $\Omega$ is a constant. Here, what we call the matter action will be a subset of the scalar field action; for example we can define $\phi_1$ to be the $\phi$ and $\phi_i$, with $i=2,\cdots, N$, to be the matter fields in $L_m$ in the previous section. 

As shown in \cite{GarciaBellido:2011de,Ferreira:2016vsc}, this system has a conserved current which is tied to the underlying Weyl symmetry of the theory.
The evolution equations for the scalar fields are
\begin{eqnarray}
\Box\phi_i-\frac{\alpha_i}{6}\phi_i R-W_{\phi_i}=0 \label{sev}
\end{eqnarray}
where $W_{\phi_i}=\partial W/\partial \phi_i$ and $R$ is the Ricci scalar which, in this
case, is given by
\begin{eqnarray}
-\frac{1}{6}\sum_{i=1}^N\alpha_i\phi_i^2R=\sum_{i=1}^N\left[(\alpha_i-1)\nabla_\mu\phi_i\nabla^\mu\phi_i
+\alpha_i\phi_i\Box\phi_i\right]+4W \nonumber \\
\end{eqnarray}
Multiplying each of the field equations \ref{sev} by $\phi_i$ and adding all of them together, one finds a
conservation law of the form $\nabla_\mu K^\mu=0$ where $K^\mu=\nabla^\mu K$ and
\begin{eqnarray}
K=\frac{1}{2}\sum_{i=1}^N(1-\alpha_i)\phi^2_i \label{ellipse}
\end{eqnarray}

We can easily understand the dynamics of this theory at the level of the background. If we take the
$\phi_i$ to be functions of time $t$ only, we have that the conservation equation give us
\begin{eqnarray}
{\ddot { K}}+3\frac{\dot a}{a}{\dot { K}}=0
\end{eqnarray}
 and can be solved to give
\begin{eqnarray}
{ K}=c_1+c_2\int\frac{dt}{a^3(t)}.
\end{eqnarray}
Therefore we find that, under general conditions, $K$ will evolve to a constant value, ${ K}\rightarrow K_0$. In other words, the
scalar fields will evolve such that their values will be constrained to lie on the  ellipse given by \ref{ellipse}.
Furthermore, one can show that, if $W_{ij}$ is non-singular, that there will be a fixed point on this ellipse where
the ratios between all possible $\phi_i^2$ are determined by the coupling constants. We then have that the
effective Planck mass, $M_{\rm Pl}$ is determined by the initial conditions of the scalar fields and the coupling
constants in the theory. This behaviour is a generalization of the simple scalar field model presented in the introduction. 

The phenomenology of the two scalar model is rich and has been extensively explored before. In particular, 
\cite{GarciaBellido:2011de,Rubio:2014wta,Bezrukov:2014ipa,Karananas:2016grc} suggested that one of the fields could be a non-minimally coupled standard model Higgs and have extensively studied the phenomenology of what they have dubbed "Higgs-Dilaton cosmology. We have explored the fixed point structure and the inflationary regime in \cite{Ferreira:2016vsc,Ferreira:2016wem} arguing that a scale-invariant, two field model can unify the IR and UV accelerated regimes into a viable cosmological model. A number of authors have explored various phenonomenological aspects of this theory in \cite{Kamada:2012se,Greenwood:2012aj,Rinaldi:2015uvu,Barrie:2016rnv}.

As before, we want to focus on what happens once the Planck mass has stabilised. In effect, the global scale invariance
of the theory will have been broken and, as one would expect, a massless Goldstone mode, the dilaton will emerge. We will show, in this case, that the dilaton is uncoupled from the matter sector. In other words, there is no fifth force. To see how this happens in practice, we change variables to
\begin{eqnarray}
\phi_i&=&e^{-\frac{\sigma}{f}} {\hat \phi}_i\nonumber \\
g_{\mu\nu}&=&e^{2\frac{\sigma}{f}} {\hat g}_{\mu\nu}
\end{eqnarray}
where ${\hat \phi}_i$ are constrained to lie on the ellipse given by
\begin{eqnarray}
{\bar K}=\frac{1}{2}\sum_{i=1}^N(1-\alpha_i){\hat \phi}^2_i=f^2
\end{eqnarray}
where $f^2$ is a constant.

Transforming the full action we find
\begin{eqnarray}
S&=&\int d^4 x\sqrt{-{\hat g}}\left[-\frac{1}{12}\sum_{i}^{N}\alpha_i{\hat \phi}_i^2
\left({\hat R}-\frac{6}{f^2}\partial_\mu\sigma\partial^\mu\sigma-\frac{6}{f^2}\Box\sigma\right)\right. \nonumber \\
& & \left.+\frac{1}{2}\sum_{i}^{N}\partial_\mu{\hat \phi}_i\partial^\mu{\hat \phi}_i +\frac{1}{2f^2}\sum_{i}^{N}{\hat \phi}_i^2\partial_\mu\sigma\partial^\mu\sigma 
+\frac{1}{f}\partial_\mu\sigma\sum_{i}^{N}{\hat \phi}_i\partial^\mu\sigma_i\right. \nonumber \\ & & \left. -W({\vec \phi})\right]  \label{msactiont1}
\end{eqnarray}
which can be integrated by parts and rewritten as
\begin{eqnarray}
S&=&\int d^4 x\sqrt{-{\hat g}}\left[-\frac{1}{12}\sum_{i}^{N}\alpha_i{\hat \phi}_i^2
{\hat R}+\frac{1}{2}\sum_{i}^{N}\partial_\mu{\hat \phi}_i\partial^\mu{\hat \phi}_i \right. \nonumber \\ & & \left.+\frac{1}{f^2}{\bar K}\partial_\mu\sigma\partial^\mu\sigma 
+\frac{1}{f}\partial_\mu\sigma\partial^\mu{\bar K}  -W({\vec \phi})\right]  \label{msactiont2}
\end{eqnarray}
Given that ${\bar K}=f^2$ is a constant we have there are no cross-terms between $\sigma$ and ${\hat \phi}_i$ and thus the dilaton is completely decoupled from everything else; in particular there are no derivative couplings between the dilaton and the remaining fields. The dilaton is canonically normalized and satisfies $\Box\sigma=0$ so that in can be set to zero in the symmetry broken phase.

It is interesting to rephrase the result in terms of ${\hat \phi}_1$ (with non-minimal coupling $\alpha\equiv\alpha_1$) and the matter action, ${\hat L}_m$ constructed from the remaining $N-1$ fields. For simplicity we restrict ourselves to $N=2$ and minimal coupling for the second field, $\chi\equiv\phi_2$). The background equations of motion fix ${\hat \phi}_1=\phi_0$ and $\chi_0=0$. We then have $f^2=\frac{1}{2}(1-\alpha)\phi^2_0$
and, as in the previous section, we can define an effective Planck mass, $M_{\rm Pl}\equiv-\frac{1}{6}\alpha\phi_0^2$. The matter action is simply
\begin{eqnarray}
{\hat L}_m=\frac{1}{2}{\hat g}^{\mu\nu}\partial_\mu\chi\partial_\nu\chi+W(\phi_0,{\hat \chi})
\end{eqnarray}
where scale invariance is now explicitly broken by the expectation value of $\phi_1$. Again, note that there is no coupling at all to the dilaton, as advertised.

\section{Adding Matter Fields.}
\label{sec:SM}

\subsection{Complex Scalar and Fermions}

We could construct a more realistic model of the matter sector which includes fermions, gauge fields and a Higgs sector. It turns out that it is sufficient to consider a fermion, $\psi$, coupled to a complex scalar field, $H$; gauge fields are conformally invariant and automatically decouple from the dilaton.
The gravitational part of the action is
\begin{eqnarray}
S_{BD}=\int d^4 x\sqrt{-g}\left[-\frac{\alpha}{12}\phi^2R+\frac{1}{2}\partial_\mu\phi\partial_\nu\phi-V(\phi)\right] \label{phi}
\end{eqnarray}
where $V=\frac{\lambda}{4}\phi^4$.

We have that fermions, $\psi$ will transform as $\psi\rightarrow \Omega^{3/2}\psi$ and therefore, the fermion action with a scale invariant mass term must take the form
%cth  fixed phi -> psi
\begin{eqnarray}
S_{\psi}=\int d^4 x\sqrt{-g}\left[\frac{i}{2}{\bar \psi}({\overrightarrow {\slashed{\nabla}}}-{\overleftarrow {\slashed{\nabla}}})\psi-g{\bar \psi}\psi_R H-g{\bar \psi}\psi_L H^*\right] \nonumber \\
\end{eqnarray}
where $\psi_L=\frac{1-\gamma^5}{2}\psi$, $\psi_R=\frac{1+\gamma^5}{2}\psi$ and $\gamma^5$ is a Dirac matrix. We have defined the covariant Dirac operator, ${\overrightarrow {\slashed{\nabla}}}=E^{a\mu}\gamma_a\partial_\mu$, where $E^{a\mu}$ is the vierbein such that $g^{\mu\nu}=\eta_{ab}E^{a\mu}E^{b\nu}$ and $\eta_{ab}$ is the Minkowski metric.

The action for the complex scalar field will take the form
%cth fixed typo upper \mu
\begin{eqnarray}
S_{H}=\int d^4 x\sqrt{-g} \left[\nabla_\mu H \nabla^\mu H^*+U(\phi,H)\right]
\end{eqnarray}
where the potential takes the form
\begin{eqnarray}
U(\phi,H)=\xi (H^*H)^2+\delta\phi^2H^*H
\end{eqnarray}
The full action is then given by $S=S_{\phi}+S_{\psi}+S_{H}$. We can easily introduce Yang-Mills gauging by suitably correcting the covariant derivative and adding in the gauge kinetic term.

Ignoring the phase of the Higgs field, we can define $H=h/\sqrt{2}$ to end up with a two scalar theory
\begin{eqnarray}
S&=&\int d^4 x\sqrt{-g}\left[\frac{1}{2}g^{\mu\nu}\partial_\mu\phi\partial_\nu\phi
+\frac{1}{2}g^{\mu\nu}\partial_\mu h\partial_\nu h \right. \nonumber \\ & & \left. +\frac{i}{2}{\bar \psi}({\overrightarrow {\slashed{\nabla}}}-{\overleftarrow {\slashed{\nabla}}})\psi-g'{\bar \psi}\psi h -W(\phi,h)-\frac{1}{2}\alpha\phi^2 R\right] \nonumber \\
\end{eqnarray}
where $g'=g/\sqrt{2}$ and $W=V+U$.

As before, we now want to extract the dilaton by transforming the fields as follows:
\begin{eqnarray}
\label{redef}
\phi&=&{\hat \phi}e^{-\frac{\sigma}{f}} \nonumber \\
h&=&{\hat h}e^{-\frac{\sigma}{f}} \nonumber \\
g_{\mu\nu}&=&e^{2\frac{\sigma}{f}}{\hat g}_{\mu\nu} \nonumber \\
E^{a\mu}&=&e^{\frac{\sigma}{f}}{\hat E}^{a\mu} \nonumber \\
\psi&=&e^{\frac{3\sigma}{2f}}\psi'
\end{eqnarray}
Applying this transformation, integrating by parts and defining the kernel, ${\bar K}=\frac{1}{2}(1-\alpha){\hat \phi}^2+\frac{1}{2}{\hat h}^2$, we find
\begin{eqnarray}
S&=&\int d^4 x\sqrt{-{\hat g}}\left[\frac{1}{2}{\hat g}^{\mu\nu}\partial_\mu{\hat \phi}\partial_\nu{\hat \phi}
+\frac{1}{2}{\hat g}^{\mu\nu}\partial_\mu{\hat h}\partial_\nu{\hat h} +\frac{1}{f}\partial_\mu\partial^\mu {\bar K}
\right. \nonumber \\ & & \left. +\frac{1}{2f^2} {\bar K}\partial_\mu\sigma\partial^\mu\sigma -\frac{1}{12}\alpha{\hat \phi}^2{\hat R} - W 
+\frac{i}{2}{\bar \psi}'({\overrightarrow {\slashed{\nabla}}}-{\overleftarrow {\slashed{\nabla}}})\psi' \right. \nonumber \\ & & \left.-g'{\bar \phi}'\phi' {\hat h}\right]
\end{eqnarray}
As in the multi-scalar case, we have a conserved Weyl current; canonically normalizing the dilaton we have
$f^2={\bar K}$ and which decouples the dilaton kinetic  term from the remaining scalar fields.

Focusing on the symmetry broken phase where we have ${\hat \phi}=\phi_0+{\tilde \phi}$ and $\phi_0\gg {\hat h}$ we can solve for $\partial{\tilde \phi}$ to get
\begin{eqnarray}
\partial{\tilde \phi}=-\frac{1}{1+\alpha}\frac{\hat h}{\phi_0}\partial {\hat h}
\end{eqnarray}
which means that $\partial{\tilde \phi}\ll \partial {\hat h}$. Furthermore, we have that ${\bar K}\simeq \frac{1}{2}(1-\alpha)\phi^2_0$ and so ${\tilde \phi}\simeq -{\hat h}^2/2(1-\alpha)$ and the leading order terms in the potential are
\begin{eqnarray}
W({\hat \phi},{\hat h})\simeq \frac{\lambda}{4}\phi_0^4+\frac{\delta'}{2}\phi_0^2{\hat h}^2+\frac{\xi'}{4}{\hat h}^4
\end{eqnarray}
where $\delta'$ and $\xi'$ can be expressed in terms of $\lambda$, $\delta$, $\xi$ and $\alpha$. The resulting action (with $M_{\rm Pl}^2=-\frac{1}{6}\phi^2_0$) is Einstein gravity:
\begin{eqnarray}
S&=&\int d^4 x\sqrt{-{\hat g}}[\frac{1}{2}M^2_{\rm Pl}{\hat R}+L_m]
\end{eqnarray}
with
\begin{eqnarray}
L_m=\frac{1}{2}{\hat g}^{\mu\nu}\partial_\mu{\hat h}\partial_\nu{\hat h} - W 
+\frac{i}{2}{\bar \psi}'({\overrightarrow {\slashed{\nabla}}}-{\overleftarrow {\slashed{\nabla}}})\psi' -g'{\bar \phi}'\phi' {\hat h} \nonumber \\
\end{eqnarray}

As in the previous cases, we have found that there is no coupling between the dilaton and the matter sector and thus, such a scale invariant theory 
won't be subject to fifth force constraints.

\subsection{Gauge Bosons}

{ Covariant (lower index) vector bosons are neutral under the Weyl 
transformations laid out in Equation \ref{redef} .
This is related to how the notion of length is contained in the covariant metric, 
and not in the coordinates under Weyl symmetry.   This means that there is a big 
difference between contravariant and covariant and
one has to be careful:  $g_{\mu \nu}\rightarrow \Omega^2g_{\mu \nu}$  
has dimensions of $L^2$ (where $L\sim$ length), but $g^{\mu \nu}\rightarrow \Omega^{-2}g^{\mu \nu}$
has dimensions of $L^{-2}$.
Hence the contravariant coordinates and their differentials,   $dx^\mu$,   
are dimensionless  numbers,  and $ ds^2= g_{\mu \nu} dx^\mu dx^\nu $ has
dimensions $L^2$ via the metric. 
Covariant coordinates,  $dx_\mu = g_{\mu \nu} dx^\nu $ thus carry $L^2$.}   

{Therefore, derivatives  $\partial_\mu  = \partial/ \partial x^\mu$
{are likewise} neutral under a Weyl transformation, $\partial_\mu \rightarrow \partial_\mu$.
When we construct a gauge covariant derivative for electromagnetism or other
unitary gauge group based theories, we introduce a vector potential and
have   $D_\mu=  \partial_\mu - ieA_\mu$. Consistency thus dictates that $A_\mu$
is also neutral under Weyl transformations, i.e. $A_\mu \rightarrow A_\mu$. 
(Note that Weyl's original gauge field enters as $D_\mu=  \partial_\mu - qe'A_\mu$, and
gauges scale transformations, where $q$ is the (mass $\sim L^{-1}$)-scale {dimension}, i.e.,
$q=1$ for $\phi$ and $q=-2$ for $g_{\mu\nu}$).}

{Hence the electromagnetic field  $ F_{\mu \nu} = \partial_\mu A_\nu -  \partial_\nu A_\mu$
is also neutral, transforming as $F_{\mu \nu}\rightarrow F_{\mu \nu} $,
but $F^{\mu \nu} = g^{\mu \rho}  g^{\nu \lambda}   F_{\rho \lambda}\rightarrow \Omega^{-4}F ^{\mu \nu}$
has dimensions of   $L^{-4}$, as an energy density.    
Since Maxwellian electromagnetic fields  $\overrightarrow{E}$ and $\overrightarrow{B}$  
have mass dimension $L^{-2}$, we see that they must be identified with  
$\overrightarrow{E}\sim F_0^i$ and $\overrightarrow{B} \sim F_i^j$ .}

{The canonical kinetic term for gauge theories is therefore}
\bea
 {\cal{L}} =-\frac{1}{4}  g^{\mu \rho}  g^{\nu \lambda}  F_{\mu \nu} F_{\rho \lambda} 
 \eea
{and we see that  that  ${\cal{L}} \rightarrow \Omega^{-4} {\cal{L}}$  is an energy density.
Since, $\sqrt{-g} \rightarrow \Omega^{4}\sqrt{-g}$,  the
action,  $S_A = \int \sqrt{-g} {\cal{ L}}$,  is invariant.
Since the Dilaton follows by replacing $\ln\Omega \rightarrow \sigma/f$, 
we see that it decouples from the classical
vector potential action.}

{What about a renormalization group running gauge coupling, $e$?  
The action can be written in the noncanonical normalization as 
$S_A= \int (1/e^2) \sqrt{-g} {\cal{ L}}$.
If we use external mass scales to
define renormalized running couplings (and for an infinitesimal 
Weyl transformation, $\Omega\simeq1+\epsilon$) we have  
$(1/e^2) \rightarrow (1/e^2) - 2( \beta(e)/e^3) \epsilon$,  (e.g. with $e^2=e^2 (\phi/M)$  
and $d \ln(\phi/M) = \epsilon $ where  $\phi \rightarrow (1+\epsilon)\phi $). }

{With ``external'' renormalization, i.e., using external
input masses $M$ to define renormalized quantities, we have the trace anomaly:}
\bea
\frac{1}{\sqrt{-g}} \frac{\delta S_A}{\delta \epsilon} & = &  - \frac{2 \beta(e)}{e^3}  {\cal{L}}  
\nonumber \\
& &\!\!\!\!\!  \!\!\!\!\!  \!\!\!\!\!  \!\!\!\!\!  \!\!\!\!\!  \!\!\!\!\!  \!\!\!\!\!  
\rightarrow \frac{\beta(e)}{2e}  F^{\mu\nu}F_{\mu\nu}  \;\;\;\makebox{(canonical normalization)}.
 \eea
{However, with ``internal'' renormalization we use  fields in the action
in place of $M$, hence
$\ln(\phi/M) \rightarrow \ln(\phi/\chi)$.
With this Weyl invariant argument of the log,
the action is invariant under $\delta{I}/{\delta \epsilon}$
and there is no trace anomaly (i.e., the associated Weyl current is conserved).
There is still running of the coupling, $g(\phi/\chi)$, but now in the variable $\ln(\phi/\chi)$.  
There is still
a physical $\Lambda_{QCD}$, but now $\Lambda_{QCD}/\chi = \exp\{-8\pi^2/|b_0| e_{QCD}^2(\chi)\}$ 
and  the ratio $\Lambda_{QCD}/\chi$ is Weyl invariant {to this order of perturbation theory}.}

{Hence, the dilaton completely decouples from gauge fields in quantum mechanics
as well, provided we use ``internal renormalization.'' This is a world in which there are
no absolute mass scales, but only dimensionless ratios of field VEV's \cite{Ferreira:2016wem},
and may be an underlying symmetry in nature.}

\subsection{Higher Dimension Operators}
\label{halos}
{ The previous discussion has been restricted to $D\leq 4$ operators. 
In fact, this provides an easy way to see why dilaton decoupling from spinors occurs: the quantity
$\partial_\mu \sigma$ is $C=+$ (i.e. charge conjugation even), while the fermionic current
${\bar \psi}\gamma_\mu \psi $ is $C=-$.
However, there will generally occur
higher dimension operators, such as those involving the nucleon (i.e. $\psi$)
that arise nonperturbatively in QCD.
For example, we might have an operator taking the form:}\
\bea
\sqrt{-g}\kappa g^{\mu\nu}\partial_\mu \sigma  \bar{\psi}\partial^\mu \psi/f\Lambda_{QCD}
\eea
%{ In a fully scale invariant theory we will have 
%$\Lambda_{QCD}^2\propto \exp(-c)\phi^2$
%and global scale invariance is maintained when $\phi =\phi_0 e^{-\sigma/f}$ (i.e., $\phi$
%might be associated with the grand-unification scale, and at one loop, 
%$c = 1/g^2(\phi)b_0 >>1$; 
We've chosen an operator that is chiral symmetry breaking 
and hence scales like $\Lambda_{QCD}/\Lambda_{QCD}^2$ 
However, the fermionic operator now has $C=+$ and the dilaton can couple
derivatively to the fermion density.

{ Now consider a compact source, like a star or planet where the nucleon
density can be approximated by a local static function $\overline{\psi }\psi
(x)=\rho(\overrightarrow{x})$ and $\psi(\overrightarrow{x})
\rightarrow 0$ for $|\overrightarrow{x}|> R$ . In this
approximation the source $\bar{\psi}\nabla_i \psi \sim (1/2)\nabla_i \rho$,
and we have a vanishing surface term:}
\bea
\int d^{3}x \; \vec{\nabla}^2\rho=0
\eea
{ If we assume approximate flat space
and we can seek a static solution for the $\sigma$ field around the source.
The equation of motion in the static limit is thus:}
\bea
\label{initial}
-\nabla ^{2}\widehat{\sigma }+\frac{\kappa }{f\Lambda_{QCD}} 
\nabla ^{2}\rho(\overrightarrow{x})=0
\eea
{ A Green's function solution
for the dilaton halo is then:}
\bea
\sigma =-\frac{\kappa }{f\Lambda_{QCD}}
\int \frac{1}{4\pi |\overrightarrow{r}-\overrightarrow{x}|}\nabla^{2}\rho(\overrightarrow{x})d^{3}x
\eea
{Performing a double integration by parts and using
$\nabla^{2}(4\pi |\overrightarrow{r}-\overrightarrow{x}|)^{-1}
=\delta^3(\overrightarrow{r}-\overrightarrow{x})$
yields:}
\bea
\sigma(\overrightarrow{r}) =-\frac{\kappa }{f\Lambda_{QCD}}\rho(\overrightarrow{r})
\eea
{This is a halo field that simply tracks the source distribution
and vanishes outside.  
Other operators and distributions might produce at most weak $1/r^3$ halos.
This is analogous to the fact that pseudoscalar fields, such
as axions, couple to 
$\bar{\Psi}\gamma^5 \Psi \sim \Psi^\dagger \overrightarrow{\sigma}\cdot
\overrightarrow{\nabla} \Psi$, where we indicate the nonrelativistic limit.
This implies that pseudoscalar fields couple to dipole densities
$\sim \overrightarrow{S}\cdot
\overrightarrow{\nabla} \rho$ (where $\overrightarrow{S}$ is a net spin
polarization). 
It is beyond the scope of the present work to determine if ultra-sensitive 
experiments could detect such a suppressed short-range halo. }

\section{Obstructions and solutions.}
\label{sec:obstructions}
We have argued that the fifth-force bounds on Brans-Dicke theories 
are absent if {a Weyl} scale invariance is only broken spontaneously. 
In the context of a complete theory of the fundamental forces this requires 
that the Standard Model (SM) should also be scale invariant with all 
masses generated spontaneously. Indeed, with the exception of the scalar potential, 
the SM Lagrangian is scale invariant and the masses of the gauge bosons, 
the quarks and the charged leptons are generated through spontaneous breaking 
of the electroweak (EW) symmetry.  

However, in the SM the spontaneous breaking of EW symmetry is triggered by the 
inclusion of a scalar mass term in the Lagrangian that explicitly breaks scale 
invariance. In the context of the SM { this term is
at the heart of the naturalness problem} that either hampers our understanding 
of the foundation of the SM or hints at new physics, depending on the eye of the beholder. 

{A rejuvenated} approach has been advocated that builds scale-invariance 
into the core of the SM \cite{Bardeen:1995kv}. The idea is that the 
spontaneous EW breaking occurs via dimensional transmutation in which 
radiative corrections drive the {Higgs ruuning} quartic scalar coupling 
negative {below} the EW scale, {leading to the Coleman-Weinberg mechanism \cite{Coleman:1973jx},
and} triggering 
spontaneous EW breaking at that scale. In the original implementations 
of this idea the classical theory is scale invariant and scale breaking 
occurs through the trace anomaly arising from the one-loop radiative 
corrections to the quadratic coupling proportional to $\ln(|H|^2/M^2)$ 
where $M$ is an explicit mass scale at which the coupling is defined and $H$ 
is the SM scalar field.  However, {introducing $M$ as an external
input mass leads to} explicit breaking of scale symmetry, {and}
such a term will induce non-derivative couplings of the  dilaton to the 
SM states, re-introducing the fifth-force bounds on the Brans-Dicke  coupling.

A more ambitious viewpoint argues that any mass scales 
that might enter via regularization and renormalization should be
{vacuum expectation values of fields in the action of the theory itself,} and thus
maintain the Weyl invariance. So, for example, logarithmic corrections 
to the action of the form $\ln(|H|^2/M^2)$ {in a scale broken theory
would} be replaced by 
$\ln(|H|^2/\phi^2)$ such that the argument of the logarithm is itself 
scale invariant. A case for this approach has been made 
in \cite{Ghilencea:2015mza,Ghilencea:2016ckm,Ferreira:2016wem}.
{This allows for nonzero $\beta$-functions and renormalization group
running of coupling constants in quantities like $\ln(|H|^2/\phi^2)$,
however the Weyl symmetry is now maintained at the quantum level.
In this case, 
scale invariance is only spontaneously broken, 
so the decoupling of the dilaton persists 
and there are no fifth-force bounds on the Brans-Dicke coupling.}

{}
There remains the question whether neutrino masses explicitly break scale symmetry. 
In the SM neutrinos are massless due to the absence of right-handed (RH) 
SM-singlet neutrinos.  If they are added to the SM then, after spontaneous 
EW breaking, neutrinos will acquire Dirac masses upon EW breaking through 
their Yukawa coupling to the SM scalar. It is possible these couplings 
{are} anomalously small and give rise to the observed neutrino masses. As for the quarks and charged leptons they do not lead to explicit scale breaking so the dilaton still decouples. 
Alternatively the LH neutrinos may acquire Majorana masses via a dimension 5 
coupling to two SM scalars through the exchange of a heavy state such a RH 
neutrino or a heavy scalar state. Provided the RH states also acquire their 
mass through spontaneous breaking of the scale symmetry the decoupling of 
the dilaton will be preserved.

\section{Discussion}
\label{sec:discussion}
In this paper we have explicitly shown that perturbatively, in scalar-tensor theories in a scale invariant universe, 
there is no fifth force. This means that the usual, extremely tight, astrophysical constraints 
can be completely evaded. We have done so by looking at a representative selection of 
actions which encapsulate the essential structure of the standard model and beyond. We 
have discussed how this result may be obstructed in the real world by explicit mass scales  
but have also {described} how to evade these obstructions.

Our result is not unexpected. We are considering a global scale symmetry which 
is spontaneously broken. From Goldstone's theorem we expect the dilaton, which is 
the mediator of the fifth force, { and to be derivatively coupled.} 
{In fact, however, the dilaton doesn't 
couple at all to the energy momentum tensor of the matter fields  in the spontaneous symmetry 
broken phase.} The dilaton obeys a damped wave equation and its dynamics are trivial: 
any residual non-zero fluctuations in the dilaton will dissipate away following the onset 
of the symmetry broken phase. The fact that the dilaton decouples from the rest of a scale 
invariant world has been alluded to before. In \cite{Fujii:1974bq}, the author constructed 
simple scalar field models involving one and two scalar fields and showed there that the 
massless scalar mode would decouple from any additional static matter sector. Our calculation 
generalizes the result of \cite{Fujii:1974bq}.

The question remains:  {does the Universe have an underlying
exact scale invariance, that is hidden by spontaneous symmetry breaking?} The conventional view,
{{\em e.g.}, such as that of string theory where scale symmetry is explicitly broken
by the string tension},
is that it is {not an exact symmetry} and, if so, our results do not hold. If there are explicit mass 
scales built into the fundamental action of the Universe then we {would be stuck with the 
extremely tight constraints on scalar-tensor theories: scalar tensor theories are
then disfavored.} But the recent resurgence in 
interest in scale-invariance, {driven in part by the discovery of a fundamental
scalar particle, the Higgs boson, is leading to a fresh look at some of the impediments and advantages
to having a scale-invariant world.} It may well {be  that scale-invariance 
solves the problems} currently facing our understanding of fundamental physics. 
If, indeed, all explicit mass scales can be dropped from our fundamental action, 
then scalar-tensor theories will be given a completely new lease on life.

\textit{Acknowledgements ---} We acknowledge discussions with D. Ghilencea,
L. Hui, A. Nicolis and T. Sotiriou. PGF acknowledges support from 
Leverhulme, STFC, BIPAC and the ERC. Part of this work was done at Fermilab, 
operated by Fermi Research Alliance, 
LLC under Contract No. DE-AC02-07CH11359 with the United States Department of Energy.

%===============================================================================
\appendix

%===============================================================================
% BIBLIOGRAPHY
%===============================================================================

\bibliographystyle{apsrev4-1}
\bibliography{STSI}

\end{document}